\documentclass[aps,prl,showpacs,amsmath,twocolumn,amssymb,superscriptaddress,letterpaper]{revtex4}
\usepackage{graphicx,color}
\usepackage{amssymb}   
\usepackage{amsmath}
\usepackage{epstopdf}
\usepackage{natbib}
\usepackage{hyperref}
\usepackage{bm}

\begin{document}
\title{Emergent Josephson current of $N=1$ chiral topological superconductor in quantum anomalous Hall insulator/superconductor heterostructures}
\author{Chui-Zhen Chen}
\affiliation{Department of Physics, Hong Kong University of Science and Technology, Clear Water Bay, Hong Kong, China. }
\author{ James Jun He}
\affiliation{Department of Physics, Hong Kong University of Science and Technology, Clear Water Bay, Hong Kong, China. }
\author{Dong-Hui Xu }
\affiliation{Department of Physics, Hubei University, Wuhan 430062, China}
\author{K. T. Law}\thanks{phlaw@ust.hk}
\affiliation{Department of Physics, Hong Kong University of Science and Technology, Clear Water Bay, Hong Kong, China. }

\begin{abstract}
Recently, a quantum anomalous Hall insulator (QAHI)/superconductor heterostructure has been realized and shows half-quantized
conductance plateaus in two-terminal conductance measurements [Q. L. He \textit{et al.}, Science {\bf357}, 294 (2017)]. The half-quantized conductance plateaus are considered as a solid evidence of chiral Majorana edge modes.
However, there is a strong debate over the origin of the half-quantized conductance plateaus. In this work, we propose a Josephson junction based on the QAHI/superconductor heterostructure to identify the existence of chiral Majorana edge modes. We find that the critical Josephson current dramatically increases to a peak value when a half-quantized conductance plateau $\sigma_{12}=e^2/2h$ is showing up for the $N=1$ chiral topological superconductor phase with a single chiral Majorana mode. Furthermore, we show that the critical Josephson current of the $N=1$ chiral topological superconductor exhibits an $h/e$-period oscillation and is robust to disorder, in contrast to the behaviors of conventional two-dimensional electron gas systems. We also estimate experimentally relevant parameters and believe that the supercurrent can be observed in experiments.
\end{abstract}

\maketitle

{\em Introduction}---
A Majorana fermion is its own antiparticle \cite{Wilczek, Kitaev1} and can appear as quasiparticle
excitations in condensed matter physics. The search for these exotic quasiparticle excitations has been one of the
central subjects in condensed matter physics, for the reason that they obey non-Abelian statistics and have potential applications in quantum computations \cite{RG,Ivanov,Fujimoto,STF,Alicea2,Kitaev2, Nayak,HasanREV,QiREV,FlensbergREV,BeenakkerREV}.
It is predicted that superconductors or superfluids with $p_x+i p_y$ paring symmetry and the $\nu=5/2$ fractional quantum Hall state can host Majorana fermions \cite{Nayak,Stone2006,Gurarie2005,RG,Moore1991}.
Majorana fermion can also exit when an $s$-wave superconductor is attached to topological insulators or semiconductors with spin-orbit coupling \cite{Sau,ORV,Fu08,Alicea2010,ZhouTong,Jia2016,Kouwenhoven}.

\begin{figure}[thp]
\centering
\includegraphics[width=3.3in]{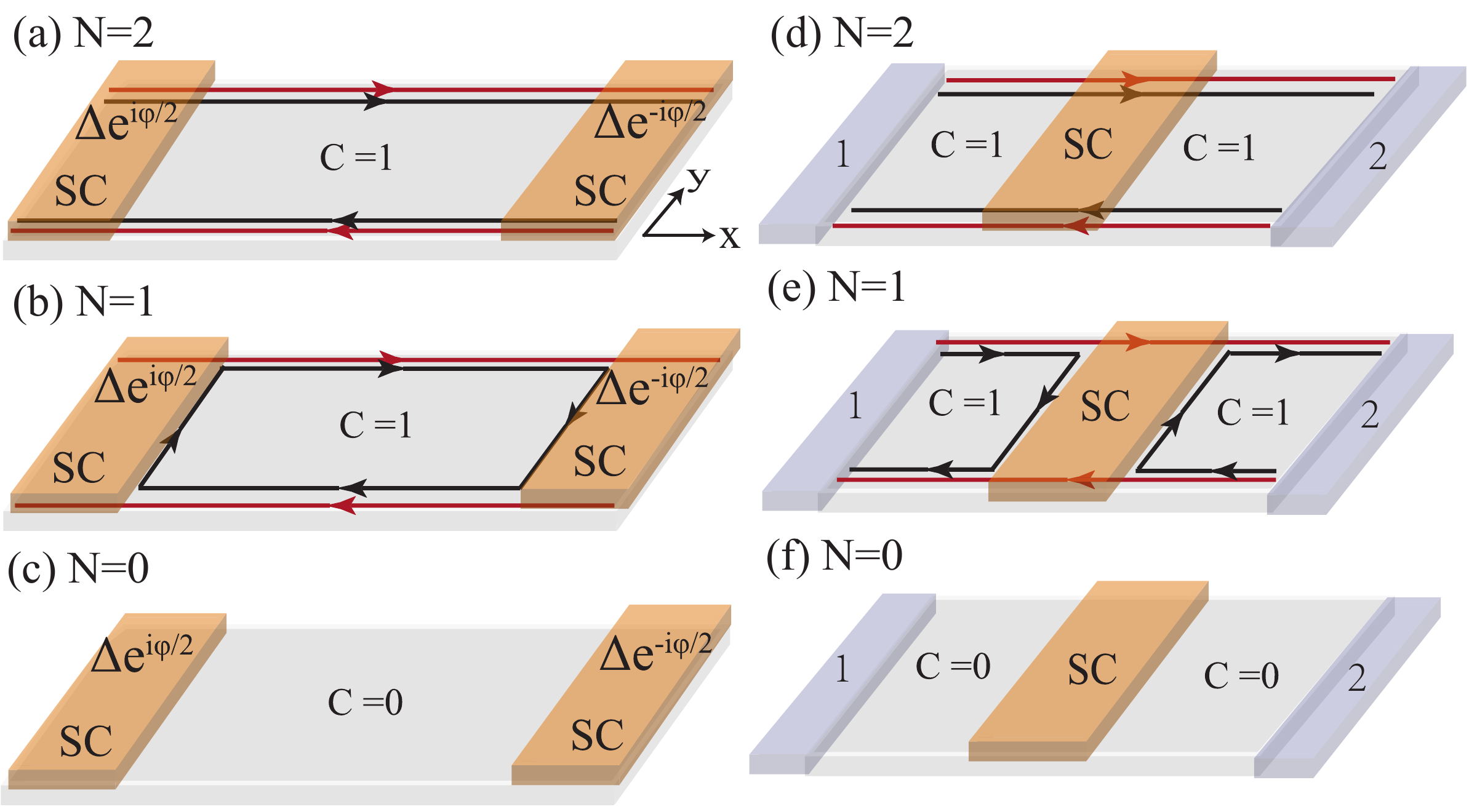}
\caption{(Color online). Schematic plot of a quantum anomalous Hall insulator (QAHI) with $C=1$ and $C=0$ under an $s$-wave superconductor
as a topological superconductor with $N=2$, $N=1$  and $N=0$ Majorana modes (black and solid lines), respectively.
(a)-(c) The Josephson junction of QAHI/superconductor heterostructure with $N=2$, $N=1$ and $N=0$, respectively. For the $N=1$ phase, the Josephson
current can be carried by a chiral Majorana mode [black line in (b)], which can exist  between two chiral topological superconductors, while there are no Andreev bound states in the junction for (a) both $N=2$  and (c) $N=0$ phases.
As a result, the critical supercurrent will dramatically increase to a peak value when QAHI/superconductor heterostructure sweeps from $N=2$ to $N=0$ phases.
(d)-(f) Two-terminal conductance measurement of QAHI/superconductor heterostructure with $N=2$, $N=1$ and $N=0$, respectively.
When the QAH/SC heterostructure is in the $N=1$ phase, an incoming fermionic mode from lead $1$ [blue cuboid in (e)] is split into two chiral Majorana modes [red and black lines in (e)], giving rise to the half-quantized conductance plateau observed in experiment \cite{Qinglin}.
\label{fig1} }
\end{figure}

Quantum anomalous Hall effect was proposed as quantum Hall effect without Landau levels by F. D. Haldane in 1988 \cite{Haldane}. Recently, the quantum anomalous Hall effect was observed \cite{Chang2013} in Cr-doped (Bi,Sb)$_2$Te$_3$ thin films experimentally as theoretically predicted previously \cite{Dai}.
When a quantum anomalous Hall insulator (QAHI) with Chern number $C=1$ 
in proximity to an $s$-wave superconductor, the system is topologically equivalent to a chiral topological superconductor (CTSC) with Chern number $N=2$, assuming that the induced superconducting paring is infinitesimal. 
For a finite superconducting gap, by tuning chemical potential or external magnetic field, a new $N = 1$ CTSC phase with a single chiral Majorana edge mode (CMEM) emerges when one of the two CMEMs in the $N = 2$ CTSC is annihilated \cite{Qi_QAHSC}. Later, theoretical studies showed that the $N = 1$ CTSC can give rise to a half-quantized longitudinal conductance plateau \cite{Chung,WangJing}.

Remarkably, half-quantized conductance plateaus (HQCPs), i.e. $\sigma_{12}=e^2/2h$, were observed by the two-terminal conductance measurements in a very recent experiment \cite{Qinglin}. The experimental setup consists of a superconductor island (Nb) on the top of QAHI (Cr-doped (Bi,Sb)$_2$Te$_3$ thin film) as shown in Fig. \ref{fig1}(e) \cite{Qinglin}. These HQCPs are regarded as a hallmark of the existence CTSCs with a single CMEM. However, there is a strong debate over the origin of the HQCPs \cite{Ji2017,Huang2017,Lian2017,Law2017,Yu-Hang2018}, because they can also possibly be attributed to trivial reasons.
For example, if Nb in the middle part is in a metallic phase instead of a superconducting phase, the system becomes a QAHI-metal-QAHI junction. Because each of the two QAHIs contributes a quantized conductance of $e^2/h$, the conductance of the QAHI-metal-QAHI junction has a half-quantized value $e^2/2h$. As a result, other evidences to verify the existence of CTSC with a single CMEM is desirable.

In this work, we propose to measure the Josephson current in a superconductor-QAHI-superconductor junction to identify the existence of CMEMs. The Josephson junction consists of a QAHI coupled to two $s$-wave superconductors on the top of two ends [see Figs. \ref{fig1}(a)-(c)]. When the QAHI/superconductor heterostructure is in the $N=2$ phase in Fig. \ref{fig1}(a), two CMEMs are not well spatially separated and the Josephson junction is equivalently connected by a chiral fermionic mode. The Josephson current is negligible, since there are no Andreev bound states between two superconductors. On the contrary, if the QAHI/superconductor heterostructure is in the $N=1$ CTSC phase in Fig. \ref{fig1}(b), the chiral fermionic mode in QAHI is spatially separated into two CMEMs at the interfaces of QAHI and CTSC [see solid black and red arrows in Fig. \ref{fig1}(b)]. In this case, the Josephson current $I_c \sim e/hE_T$ is carried by the CMEM [black line in Fig. \ref{fig1}(b)] and decays as $1/L$, with the Thouless energy $E_T\propto1/L$ and the circumference $L$.
For comparison, a similar setup measuring two-terminal conductance of a QAHI/superconductor heterostructure is shown in Figs. \ref{fig1}(d)-(f) where an $s$-wave superconductor is placed on the top of the central region of a QAHI. HQCPs emerge in this setup when $N=\pm1$ CTSC phases are realized in the central region [see Fig. \ref{fig1}(e)].

In the following, we first introduce the model Hamiltonian for the QAHI/superconductor heterostructure, and then
calculate the Josephson current as well as the two-terminal conductance by using the recursive Green's function method.
Importantly, we find two critical Josephson current peaks when the two HQCPs show up at $N=\pm1$ CTSC phases, respectively.
This provides important transport features to identify the existence of a single CMEM. At last, we show that the critical Josephson current exhibits an $h/e$-period oscillation and is robust to disorder, in contrast to the behaviors of conventional two-dimensional electron gas systems.
We also estimate experimentally relevant parameters and believe that the Josephson current can be observed experimentally.

{\em Model Hamiltonians}---
The effective Hamiltonian of  Cr-doped (Bi,Sb)$_2$Te$_3$ thin film can be written as \cite{Dai,WangJing3}
\begin{eqnarray}
\!\!\!\!\!\!\!\!\!\!\!
H \!&=&\!\hbar v_F (k_y\sigma_x\tau_z \!-\!k_x \sigma_y\tau_z) \!+\!  m({\bf k}) \tau_x \!+\!  M'_z \sigma_z \!+\! V({\bf r})
\end{eqnarray}
in the basis of a four-component electron operator $\Psi_{\bf k}=[\psi_{{\bf k}t \uparrow}, \psi_{{\bf k}t \downarrow},\psi_{{\bf k}b \uparrow},\psi_{{\bf k}b\downarrow}]$
with the momentum ${\bf k}$. Here $t$ ($b$) denotes the top (bottom) layer of topological insulator surfaces and $\uparrow$ ($\downarrow$) is the spin index.
The Pauli matrices $\sigma_{x,y,z}$ and $\tau_{x,z}$ are defined in spin and layer spaces, respectively.
The onsite disorder is $V({\bf r})=diag\{V_{t,\uparrow},V_{t,\downarrow},V_{b,\uparrow},V_{b,\downarrow}\}$ with $V_{t/b,\uparrow/\downarrow}$ uniformly distributed in $[-W/2,W/2]$ and the disorder strength $W$.
$M'_z = M_z + M_0$ represents the spin splitting in the $z$ direction \cite{WangJing3}, where $M_0$ is the spin splitting without magnetic field $B$ and $M_z=\mu_M B$ with a proportionality constant $\mu_M$.
$m({\bf k}) = m_0 - m_1 k^2$ describes the effective coupling between the top layer and the bottom layer.
In the clean limit, the system is a QAHI with Chern number $C=sgn(M'_z)$ if $|M'_z|>|m_0|$, while it becomes a normal insulator with  zero Chern number when $|M'_z|<|m_0|$.
In our numerical simulations, we discretize the model Hamiltonian into a tight-binding model 
and add the Peirls substitution $\phi_{n,m}=\frac{e}{\hbar}\int^{n}_{m}{\bf A\cdot}d{\bf r}$ to the hopping term between ${\bf r}_n$ and ${\bf r}_m$ lattice sites. We choose Landau gauge for the vector potential ${\bf A}=(-By,0)$ with the coordinate $y$. The Fermi velocity is $v_F=1$, $m_1=1$, $m_0 = -0.03$, $M_0=0.17$ and $\mu_M= 10^{4}$.

\begin{figure}[bht]
\centering
\includegraphics[width=3.5in]{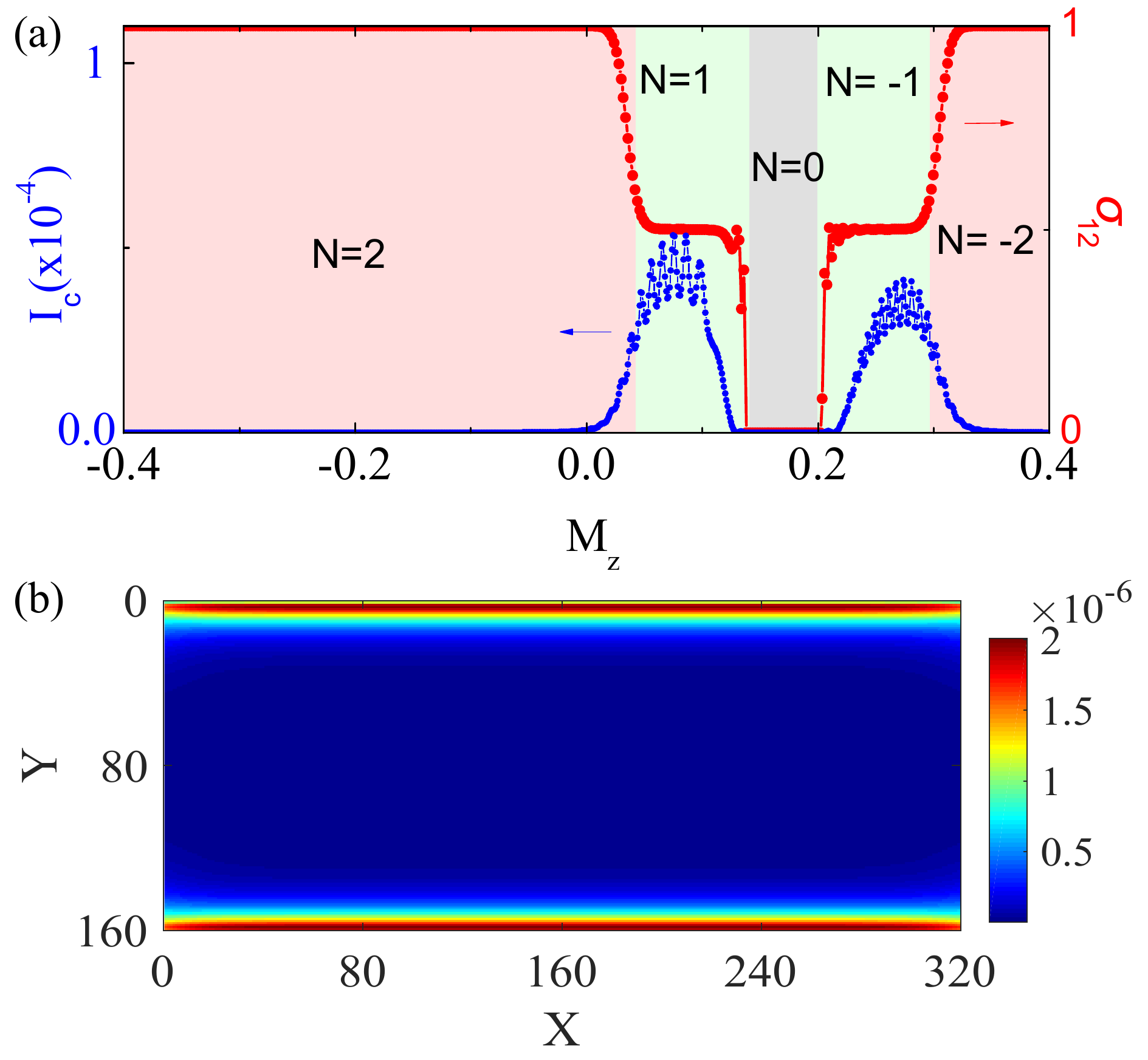}
\caption{(Color online).
(a) A comparative plot of the critical Josephson current $I_c$ and two-terminal conductance $\sigma_{12}$ versus the magnetization $M_z$.
During the magnetic field scanning, the QAHI/supercondutor heterostructure shows a various of topological phases with Chern number $N$.
It is oblivious that there is a $I_c$ peak for each of the two half quantized plateaus $\sigma_{12}=e^2/2h$, originating from $N=\pm1$ chiral topological superconductor phases, respectively.
(b) The Josephson current distribution for $N=1$ chiral topological superconductor with $M_z=0.07$ and size $W\times L=160\times320$.
\label{fig2} }
\end{figure}
In proximity to an $s$-wave superconductor, the Bogoliubov-de Gennes (BdG) Hamiltonian of the QAHI/superconductor heterostructure is given by
\begin{eqnarray}
  H_{\text{BdG}} &=& \left(
                \begin{array}{cc}
                  H(\mathbf{k})-\mu & \Delta \\
                  \Delta^{\dagger} & -H^{\ast}(\mathbf{-k}) +\mu \\
                \end{array}
              \right)
   \\
  \Delta &=& \left(
               \begin{array}{cc}
                 \Delta_{t}i\sigma_y & 0 \\
                 0 & \Delta_{b}i\sigma_y \\
               \end{array}
             \right) \exp(i\Phi_s).  \nonumber
\end{eqnarray}
in the basis of $\Phi_{\bf k} = [\Psi_{\bf k},\Psi^{\dagger}_{\bf -k}]^{T}$.
Here $\Delta_{t}=0.12$ and $\Delta_{b}=0$ are the induced pairing potentials on the top and bottom layers, respectively.
$\Phi_s$ is the phase factor of $\Delta$ and the chemical potential $\mu=0$ .

{\em Emergent Josephson current in chiral TSC}---
In proximity to an $s$-wave superconductor, a $C=1$ QAH state can regarded as an $N=2$ CTSC state and it can show two new phases ($N=\pm1$) with a single CMEM by tuning the magnetization $M_z$, as shown by the green color region in Fig. \ref{fig2}(a).
When the condition $(-\sqrt{\Delta_t^2+4m_0^2}-\Delta_t)/2<M'_z<(-\sqrt{\Delta_t^2+4m_0^2}+\Delta_t$)/2 is satisfied \cite{Qi_QAHSC, Chung, WangJing}, the system enters the $N=1$ CTSC phase. Since a $C=1$ QAHI is topologically equivalent to an $N=2$ CTSC phase, there must be a CMEM at the boundary between the QAHI and the $N=1$ CTSC as depicted in
Fig. \ref{fig1}(b). As a result, when a chiral fermionic mode in the QAH is injecting into the $N=1$ CTSC, it splits into two branches of CMEMs. One branch of the CMEMs tunneling into the $N=1$ CTSC, the other branch of the CMEMs is trapped as Andreev bound states between the two superconductors [see black line in Fig. \ref{fig1}(b)]. These Andreev bound states will give rise to Josephson current when two superconductors phases are different. Similarly, in Fig. \ref{fig1}(e), a chiral fermionic mode splits into two CMEMs at the boundary between the QAHI and the $N=1$ CTSC. One CMEM tunnels through the $N=1$ CTSC phase while the other is reflected, giving rise to a HQCP \cite{Chung,WangJing}.

Next, we numerically evaluated the Josephson current and the two-terminal conductance by the recursive Green's function method \cite{Datta1996,Furusaki,PatrickGF}. The main results are shown in Fig. \ref{fig2}(a). We can see that the critical Josephson current $I_c$ has a peak value for each of two HQCPs $\sigma_{12}=e^2/2h$. This is the central conclusion of this work. The Josephson current is concentrated on the two edges of the sample when the system is in the $N=1$ CTSC phase with $M_z = 0.07$ [see Fig. \ref{fig2}(b)]. For the $N=2$ and $N=0$ CTSCs, because there are no Andreev bound states between two superconductors, the critical Josephson current is almost zero. On the other hand, two $N=1$ superconductors will trap Andreev bound states in the QAH region, resulting in critical Josephson current $I_c$ peak we discussed above. Furthermore, we find that the critical Josephson current $I_c$ peaks are strongly oscillated, due to periodic change of the magnetic flux in the QAH region.

\begin{figure}[tbh]
\centering
\includegraphics[width=3.3in]{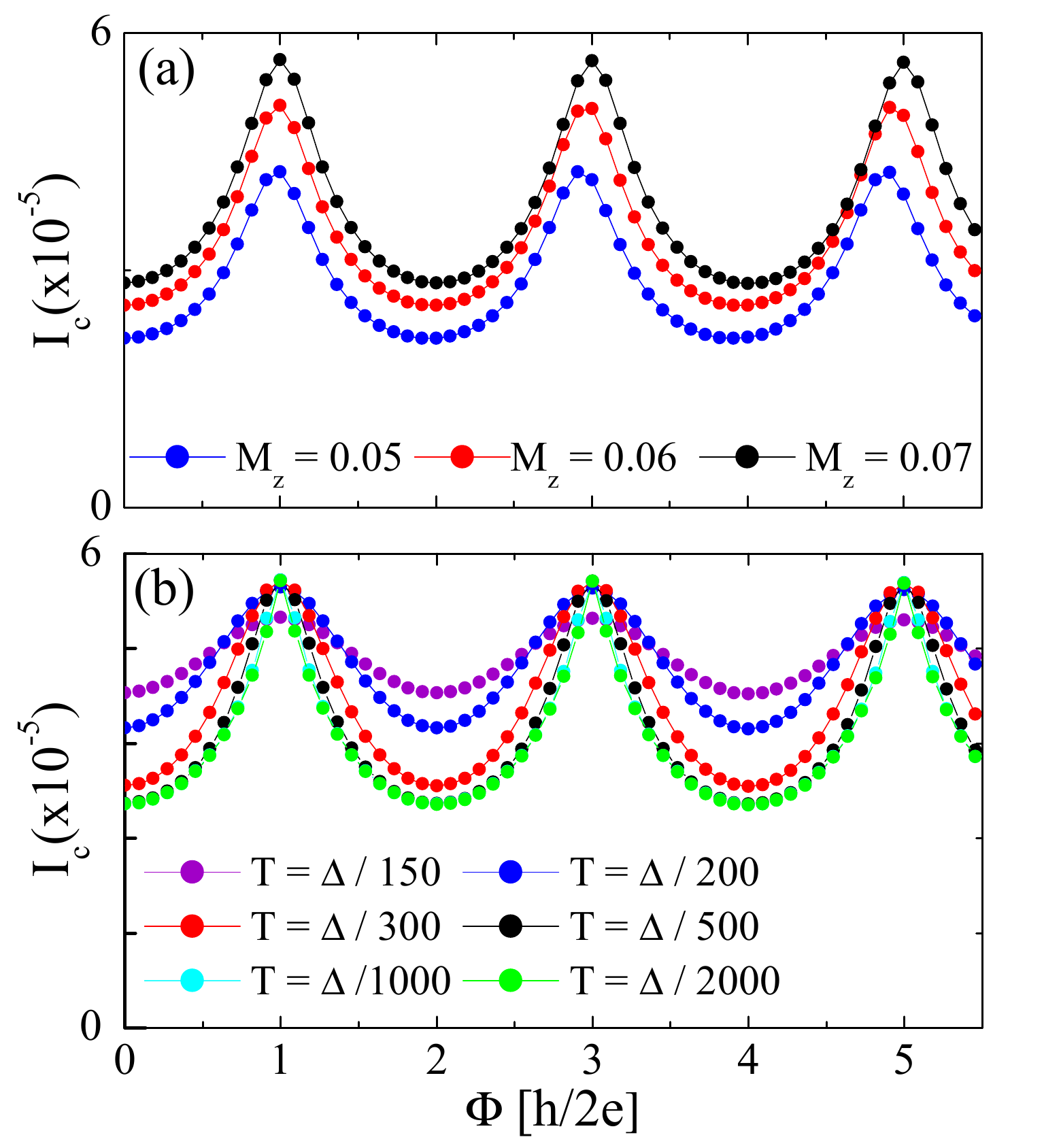}
\caption{(Color online). The critical Josephson current $I_{c}$ with a period of $h/e$ versus the magnetic flux $\Phi[h/2e]$ for different magnetizations (a) $M_z=0.05$ to $0.07$ at the temperature $T=\Delta/500$; and for different temperatures (b) $T=\Delta/150$ to $\Delta/2000$ at $M_z=0.07$.
\label{fig3} }
\end{figure}

Now let's come to investigate the oscillation of the Josephson current in the $N=1$ CTSC phase.
In Fig. \ref{fig3}, the critical Josephson current $I_c$ oscillates with a period of $h/e$ as a function of the magnetic flux $\Phi$ for different magnetizations $M_z$ \cite{note1}. This is consistent with previous results of the three-dimensional topological insulator based Josephson junction \cite{Shapiro}.
Since the chiral edge mode has a $\pi$ Berry phase, the $I_c$ has a peak value at $\Phi=h/2e$ instead of $\Phi=0$.
Therefore, the $h/e$-period oscillation of Josephson current is the key feature of the one-dimensional chiral edge mode \cite{Zyuzin,Ma,Shapiro,Liu2017}, which was also discovered in the quantum Hall insulator based Josephson junction \cite{Zyuzin,Ma,Liu2017}, and is very distinct from the behaviors of conventional two-dimensional electron gas systems.
Moreover, upon decreasing the temperature from $T=\Delta/200$ to $T=\Delta/2000$, we find that the magnitude of current oscillation becomes more pronounced and hardly varies with temperature after $T=\Delta/1000=1.2\times10^{-4}$,
because the Thouless energy $E_T\approx10^{-3}\gg T$.

\begin{figure}[tbh]
\centering
\includegraphics[width=3.5in]{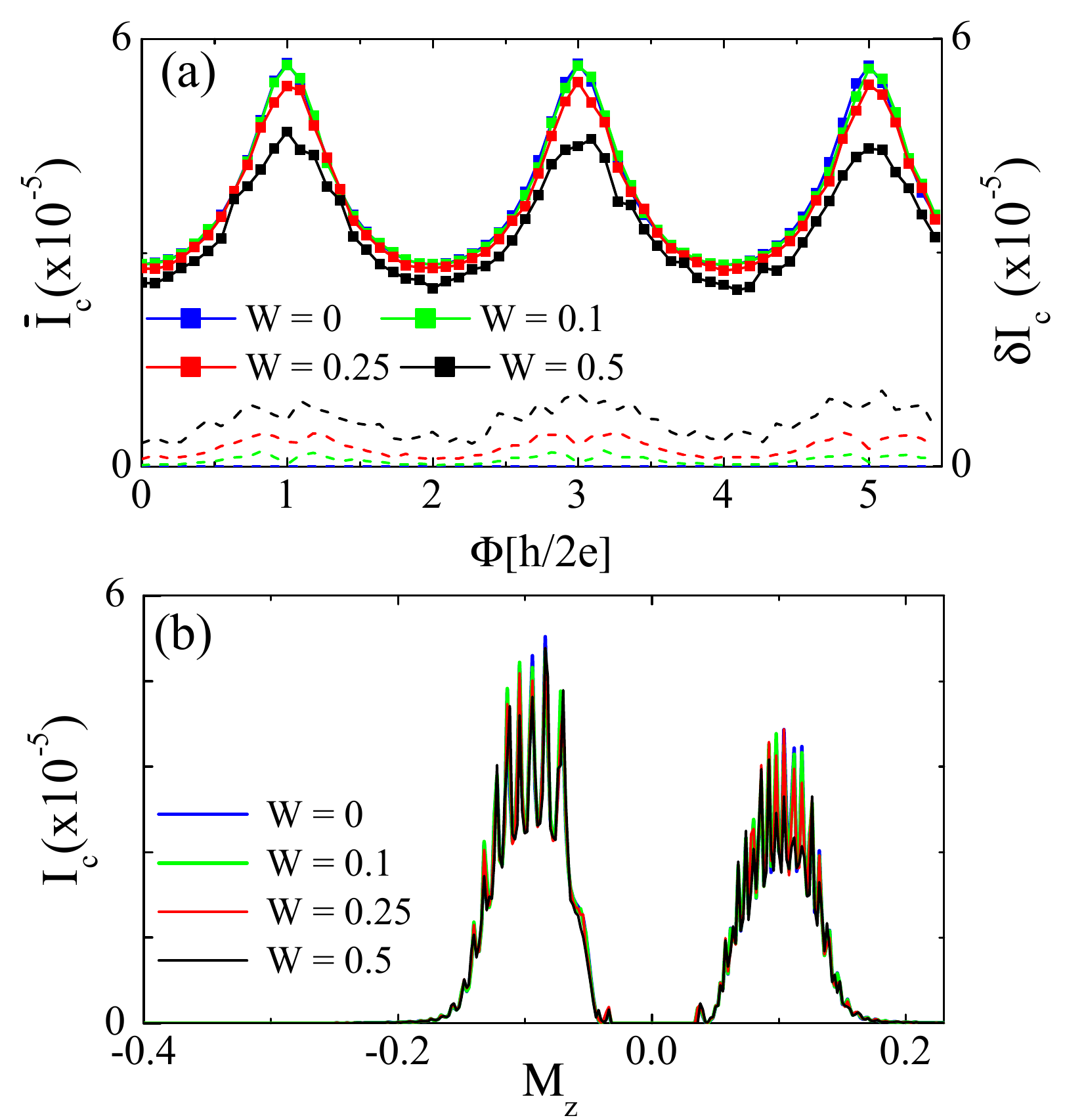}
\caption{(Color online). (a) Plots of the disorder-averaged critical Josephson current $\bar{I}_{c}$ and its fluctuation $\delta I_{c}$ (dash lines) as a function of the magnetic phase flux $\Phi[h/2e]$ at $M_z=0.07$ and the temperature $T=\Delta/500$. (b) The critical Josephson current $I_{c}$ versus the magnetization $M_z$ for various disorder strength. Other model parameters are the same as Fig. \ref{fig2}(a). We average over 100 disorder configurations in (a) and one configuration in (b).
\label{fig4} }
\end{figure}

Since the QAHI is achieved by doping topological insulator thin films with magnetic impurities, we study disorder effects on the critical Josephson current in the following. We plot the disorder-averaged critical Josephson
current  $\bar{I}_c$ in Fig. \ref{fig4}. In Fig. \ref{fig4}.(a), $\bar{I}_c$ remains stable when $W=0.1$ due to the fact that the chiral edge mode is robust to moderate disorder. We can see that the current fluctuation $\delta I_c$ becomes comparable to the amplitude of $h/e$ periodic oscillation for $W=0.5$ [see Fig. \ref{fig4}(a)]. This means that the $h/e$-period oscillation starts to be destroyed, because the chiral propagating edge mode in QAH region can be scattered to the opposite side assisted by the disorder induced bulk states. However, the critical Josephson current $I_c$ peak is very sustainable and clear in Fig. \ref{fig4}(b) even at $W=0.5$, as long as the Andreev bound states remain between two superconductors. Therefore, we conclude that the critical Josephson current $I_c$ peak is more stable than $h/e$-period oscillation, and they are both robust to the moderate disorder because the chiral edge states are topologically protected.

The critical Josephson current can be estimated by $I_c\approx \frac{e\Gamma^2}{\hbar} E_T=\frac{e\Gamma^2}{\hbar} \frac{\hbar v_F}{2L}$ with the Thouless energy $E_T$ ,
interface transparency $\Gamma$, the Fermi velocity $v_F$ and the circumference $L$ of the CMEM trapped between two superconductors \cite{Bardeen1972}. When $L=1 -2\mu m$, the critical Josephson current is $I_c=50 - 100$nA with $v_F=5\times10^5 m/s$ \cite{Xue2010} and $\Gamma=0.5$. Therefore, we believe that the critical Josephson current $I_c$ peak can be observed in experiments.

{\em Discussion and Conclusion}---
In the above analysis, we show that the critical Josephson current $I_c$ has a robust peak value for each of the two HQCPs $\sigma_{12}=e^2/2h$ due to the existence of the single CMEM in the $N=\pm1$ CTSCs.
If the $N=1$ CTSC in Fig. \ref{fig1}(e) is so strongly disordered that it becomes a gapless Majorana metal, and the gapless Majorana metal can still lead to a HQCP \cite{Huang2017}. However, in this circumstance, the Josephson junction in Fig. \ref{fig1}(b) can not sustain a Josephson current peak. That's because the Andreev bound states cannot be trapped between two gapless Majorana metals and thus the Josephson current is extremely small and strongly fluctuated. In conclusion, we find that a comparative study of the Josephson current and the two-terminal conductance can provide a more reliable evidences for the existence of single CMEM in the QAHI/superconductor heterostructure.

After submitting this paper to Physical Review B in July 2018, we noticed a study of a similar experimental setup in arXiv:1810.01891 \cite{Chang-An2018}.

{\emph{Acknowledgement.}}--- We thank Kang L. Wang, Qing Lin He and Jie Liu for illuminating discussions. The authors acknowledge the support of HKRGC through C6026-16W, the Croucher Foundation and the Tai-chin Lo Foundation. D.-H.X. is supported by the National Natural Science Foundation of China (Grant No. 11704106), the Scientific Research Project of Education Department of Hubei Province (Grant No. Q20171005) and the Chutian Scholars Program in Hubei Province.


\begin{thebibliography}{99}
\bibitem{Wilczek} F. Wilczek, Nat. Phys. {\bf 5}, 614 (2009).
\bibitem{Kitaev1} A. Y. Kitaev, Physics-Uspekhi {\bf44}, 131 (2001).
\bibitem{RG} N. Read and D. Green, Phys. Rev. B {\bf61}, 10267 (2000).
\bibitem{Ivanov} D. A. Ivanov, Phys. Rev. Lett. {\bf86}, 268 (2001).
\bibitem{Fujimoto} S. Fujimoto, Phys. Rev. B {\bf77}, 220501 (2008).
\bibitem{STF} M. Sato, Y. Takahashi, S. Fujimoto, Phys. Rev. Lett. {\bf103} 020401 (2009).
\bibitem{Alicea2} J. Alicea, Y. Oreg, G. Refael, F. von Oppen, M. P. A. Fisher Nat. Phys. {\bf7},1915 (2011).
\bibitem{Kitaev2} A. Kitaev, Ann. Phys. {\bf 303}, 2-30 (2003).
\bibitem{HasanREV} M. Z. Hasan and C. L. Kane, Rev. Mod. Phys. {\bf 82}, 3045 (2010).
\bibitem{QiREV} X. L. Qi and S. C. Zhang, Rev. Mod. Phys. {\bf 83}, 1057 (2011)
\bibitem{FlensbergREV} M. Leijnse and K. Flensberg, Semicond. Sci. Technol. {\bf 27}, 124003 (2012).
\bibitem{BeenakkerREV} C. W. J. Beenakker, Annu. Rev. Condens. Matter Phys. {\bf 4}, 113 (2013).

\bibitem{Nayak} C. Nayak, S. H. Simon, A. Stern, M. S. Freedman and S. Das Sarma,  Rev. Mod. Phys. {\bf 80}, 1083-1159 (2008).
\bibitem{Moore1991} G. Moore and N. Read, Nucl. Phys. B {\bf 360 }, 362 (1991).
\bibitem{Gurarie2005} V. Gurarie, L. Radzihovsky, and A.V. Andreev, Phys. Rev. Lett. {\bf 94}, 230403 (2005).
\bibitem{Stone2006} M. Stone and S. B. Chung, Phys. Rev. B {\bf 73 }, 014505 (2006).


\bibitem{Sau} J. D. Sau, R. M. Lutchyn, S. Tewari and S. Das Sarma, Phys. Rev. Lett. {\bf 104}, 040502 (2010).
\bibitem{ORV} Y. Oreg, G. Refael and F. von Oppen, Phys. Rev. Lett. {\bf 105}, 177002 (2010).
\bibitem{Fu08} L. Fu and C. L. Kane, Phys. Rev. Lett. {\bf 100}, 096407 (2008).
\bibitem{Kouwenhoven}V. Mourik, K. Zuo, S. M. Frolov, S. R. Plissard, E. P. A. M. Bakkers, and L. P. Kouwenhoven, Science {\bf 336}, 1003 (2012).
\bibitem{ZhouTong} B. T. Zhou, N. F. Q. Yuan, H. L. Jiang and K. T. Law, Phys. Rev. B {\bf 93}, 180501(R).
\bibitem{Jia2016} Hao-Hua Sun et al., Phys. Rev. Lett. {\bf 116}, 257003 (2016).
\bibitem{Alicea2010} J. Alicea, Phys. Rev. B {\bf 81}, 125318 (2010).

\bibitem{Haldane} F. D. M. Haldane, Phys. Rev. Lett. {\bf 61} 2015 (1988).
\bibitem{Chang2013}C.-Z. Chang, J. Zhang, X. Feng, J. Shen, Z. Zhang, M. Guo, K. Li, Y. Ou, P. Wei, L.-L. Wang, Z.-Q. Ji, Y. Feng, S. Ji, X. Chen, J. Jia, X. Dai, Z. Fang, S.-C. Zhang, K. He, Y. Wang, L. Lu, X.-C. Ma, and Q.-K. Xue, Science 340, {\bf 167} (2013).
\bibitem{Dai} R. Yu, W. Zhang, H. J. Zhang, S. C. Zhang, X. Dai and Z. Fang, Science {\bf 329}, 61 (2010).


\bibitem{Qi_QAHSC} X. L. Qi, T. L. Hughes and S. C. Zhang, Phys. Rev. B {\bf 82}, 184516 (2010).
\bibitem{Chung} S. B. Chung, X. L. Qi, J. Maciejko and S. C. Zhang, Phys. Rev. B {\bf 83}, 100512(R) (2011).
\bibitem{WangJing} J. Wang, Q. Zhou, B. Lian and S. C. Zhang, Phys. Rev. B {\bf 92}, 064520 (2015).



\bibitem{Qinglin}  Q. L. He, L. Pan, A. L. Stern, E. Burks, X. Che, G. Yin, J. Wang, B. Lian, Q. Zhou, E. S. Choi, K. Murata, X. Kou, T. Nie, Q. Shao, Y. Fan, S.-C. Zhang, K. Liu, J.
Xia, and K. L. Wang, Science {\bf357}, 294 (2017).
\bibitem{Law2017} C.-Z. Chen, J. J. He, D.-H. Xu, and K. T. Law, Phys. Rev. B {\bf96}, 041118(R) (2017).
\bibitem{Ji2017}W. Ji and X.-G. Wen, Phys. Rev. Lett. {\bf 120}, 107002 (2018).
\bibitem{Huang2017}Y. Huang, F. Setiawan and J. D. Sau,  Phys. Rev. B {\bf 97}, 100501(R) (2018).
\bibitem{Lian2017}B. Lian, J. Wang, X.-Q. Sun, A. Vaezi, and S.-C. Zhang, Phys. Rev. B {\bf 97}, 125408 (2018).
\bibitem{Yu-Hang2018}Yu-Hang Li, Jie Liu, Haiwen Liu, Hua Jiang, Qing-Feng Sun, and X. C. Xie, Phys. Rev. B {\bf98}, 045141(2018).
\bibitem{WangJing3} J. Wang, B. Lian and S. C. Zhang, Phys. Rev. B {\bf 89}, 085106 (2014).

\bibitem{Furusaki} A. Furusaki, Physica (Amsterdam) {\bf 203B}, 214 (1994); Y. Asano, Phys. Rev. B {\bf63}, 052512 (2001).

\bibitem{Datta1996} M. P. Anantram and S. Datta, Phys.Rev.B {\bf53}, 16390 (1996).
\bibitem{PatrickGF}  P. A. Lee and D. S. Fisher, Phys. Rev. Lett. {\bf 47}, 882 (1981); D. S. Fisher and P. A. Lee, Phys. Rev. B {\bf23}, 6851 (1981).


\bibitem{note1} The magnetic flux $\Phi=A_{eff} \times B$, where $B$ is magnetic field  and $A_{eff}=L\times W/1.1$ is effective area with the length $L$ and the width $W$.
We define the area $A_{eff}$ in such way to include the finite broadening of one dimensional chiral edge states.
\bibitem{Shapiro}D. S. Shapiro, A. Shnirman, and A. D. Mirlin, Phys. Rev. B {\bf93}, 155411 (2016).
\bibitem{Ma} M. Ma and A. Yu. Zyuzin, Europhys. Lett. {\bf21}, 941 (1993).
\bibitem{Zyuzin}A. Yu. Zyuzin, Phys. Rev. B {\bf50}, 323 (1994).
\bibitem{Liu2017} J. Liu, H. Liu, J. Song, Q.-F. Sun, and X. C. Xie, Phys. Rev. B {\bf96}, 045401 (2017).


\bibitem{Bardeen1972} C. Ishii, Prog. Theor. Phys. 44, 1525 (1970); J. Bardeen and J. L. Johnson, Phys. Rev. B 5, 72 (1972) ;
A. V. Svidzinsky, T. N. Antsygina, and E. N. Bratus, J. Low Temp. Phys. 10, 131 (1973).
\bibitem{Xue2010}Y. Zhang, K. He, C.-Z. Chang, C.-L. Song, L.-L. Wang, X. Chen, J.-F. Jia, Z. Fang, X. Dai, W.-Y. Shan,
 S.-Q. Shen, Q. Niu, X.-L. Qi, S.-C. Zhang, X.-C. Ma, and Q.-K. Xue, Nat. Phys. {\bf6}, 584 (2010).
\bibitem{Chang-An2018}Chang-An Li, Jian Li, and Shun-Qing Shen, arXiv:1810.01891.
\end{thebibliography}
\end{document}